\documentclass[aps,prd,preprintnumbers]{revtex4}
\usepackage{graphicx}                              
\usepackage{amsmath,amsfonts}
\graphicspath{{./pictures/}}                       

\usepackage{epsfig}
\usepackage{amssymb}
\usepackage{amsfonts}
\usepackage{float}
\usepackage{bbm}
\usepackage{psfrag}
\newcommand{\vs}{\vspace}
\newcommand{\hs}{\hspace}

\newcommand{\bdm}{\begin{displaymath}}
\newcommand{\edm}{\end{displaymath}}
\newcommand{\beq}{\begin{equation}}
\newcommand{\eeq}{\end{equation}}
\newcommand{\bea}{\begin{eqnarray}}
\newcommand{\eea}{\end{eqnarray}}
\newcommand{\bit}{\begin{itemize}}
\newcommand{\eit}{\end{itemize}}
\newcommand{\bc}{\begin{center}}
\newcommand{\ec}{\end{center}}
\newcommand{\re}{\relax{\rm I\kern-.18em R}}
\newcommand{\ID}{\mathbbm{1}}

\newcommand{\fhs}[1]{\mbox{\hs{#1}}}
\newcommand{\ie}{{\it i.e.} }
\newcommand{\Dov}{{\cal D}^{(ov)}}
\newcommand{\D}{\Dov}

\newcommand{\ImpSpace}{{\cal P}}

\newcommand{\sumFL}{\sum\limits_{i=1}^{N_f}}

\newcommand{\fermiMat}{{\cal M}}

\begin{document}
\preprint{HU-EP-08/64, DESY 08-191}

\title{Lower Higgs boson mass bounds from a chirally invariant lattice Higgs-Yukawa model with overlap fermions} 

\author{P. Gerhold$^{a,b}$, K. Jansen$^b$}
\affiliation{$^a$Humboldt-Universit\"at zu Berlin, Institut f\"ur Physik, 
Newtonstr. 15, D-12489 Berlin, Germany\\
$^b$DESY,\\
 Platanenallee 6, D-15738 Zeuthen, Germany}

\date{February 24, 2009}

\begin{abstract}
We study the coupling parameter dependence of the Higgs boson mass in a chirally invariant lattice
Higgs-Yukawa model emulating the same Yukawa coupling structure as in the Higgs-fermion sector
of the Standard Model. Eventually, the aim is to establish non-perturbative upper and lower
Higgs boson mass bounds derived from first principles, in particular not
relying on vacuum stability considerations for the latter case. 
Here, we present our lattice results for the lower Higgs boson mass bound at several 
values of the cutoff $\Lambda$ and compare them to corresponding analytical calculations 
based on the effective potential as obtained from lattice perturbation theory. Furthermore, 
we give a brief outlook towards the calculation of the upper Higgs boson mass bound.
\end{abstract}


\maketitle

\section{Introduction}
\label{sec:Introduction}

With the existing evidence for the triviality of the Higgs sector~\cite{Aizenman:1981du,Frohlich:1982tw,Luscher:1988uq, Hasenfratz:1987eh, Kuti:1987nr, Hasenfratz:1988kr,Gockeler:1992zj} 
of the Standard Model, rendering the removal 
of the cutoff $\Lambda$ from the theory impossible, physical quantities in this sector will, in general, 
depend on the cutoff. Though this restriction strongly limits the predictive power of any calculation 
performed in the Higgs sector, it opens up the possibility of drawing conclusions on the energy 
scale $\Lambda$ at which new physics has to set in, once, for example, the Higgs boson mass has been 
determined experimentally.

Besides the obvious interest in narrowing the energy interval of possible Higgs boson masses consistent with
phenomenology, the latter observation was the main motivation for the great efforts 
spent on the determination of cutoff-dependent upper and lower Higgs boson mass bounds. 
In perturbation theory such bounds have been derived from the criterion of the Landau pole being situated
beyond the cutoff of the theory~(see e.g.~\cite{Cabibbo:1979ay, Dashen:1983ts, Lindner:1985uk}), 
from unitarity requirements~(see e.g.~\cite{Dicus:1992vj, Lee:1977eg, Marciano:1989ns})
and from vacuum stability considerations~(see e.g.~\cite{Cabibbo:1979ay, Linde:1975sw,Weinberg:1976pe,Linde:1977mm, Sher:1988mj, Lindner:1988ww}),
as reviewed in~\cite{Hagiwara:2002fs}.

However, the arguments relying on the vacuum instability have to be considered with care. It is
well known that the constraint effective potential, which becomes equal to the effective potential in the
thermodynamic limit, is convex and can therefore never be unstable~\cite{O'Raifeartaigh:1986hi}. Additionally, it has recently
been argued that the appearance of vacuum instability is actually caused by the incorrect removal of the
cutoff from the trivial theory~\cite{Holland:2003jr,Holland:2004sd}. In this work, we will not address 
the question of the validity of the vacuum instability argument but rather investigate whether the lower mass
constraint can also be established without relying on vacuum stability considerations.
Additionally, the validity of the perturbatively obtained upper Higgs boson mass bounds is unclear, because
the corresponding perturbative calculations had to be performed at rather large values
of the renormalized quartic coupling constant. The latter two remarks make the Higgs boson mass bound
determination an interesting subject for non-perturbative investigations starting from first
principles, such as the lattice approach. 

The main objective of lattice studies of the pure Higgs and Higgs-Yukawa sector of the electroweak Standard Model 
has therefore been the non-perturbative determination of the cutoff-dependence
of the upper and lower Higgs boson mass bounds~\cite{Hasenfratz:1987eh, Kuti:1987nr, Hasenfratz:1988kr,Bhanot:1990ai,Holland:2003jr,Holland:2004sd} 
as well as its decay properties~\cite{Gockeler:1994rx}. There are two main developments that 
warrant the reconsideration of these questions. First, with the advent of the LHC, 
we are to expect that properties of the Standard Model Higgs boson, such as 
the mass and the decay width, will be revealed experimentally. Second, there 
is, in contrast to the situation of earlier investigations of lattice 
Higgs-Yukawa  models~\cite{Smit:1989tz,Shigemitsu:1991tc,Golterman:1990nx,book:Jersak,book:Montvay,Golterman:1992ye,Jansen:1994ym}, 
which suffered from their inability to restore chiral symmetry in the
continuum limit while lifting the unwanted fermion doublers at the same
time, a consistent formulation of a Higgs-Yukawa model with an exact 
lattice chiral symmetry~\cite{Luscher:1998pq} based on the Ginsparg-Wilson 
relation~\cite{Ginsparg:1981bj}. This new development allows to maintain the chiral character of the 
Higgs-fermion coupling structure of the Standard Model on the lattice
while simultaneously lifting the fermion doublers, thus eliminating manifestly the main objection to the earlier 
investigations. The interest in lattice Higgs-Yukawa models has therefore recently been
renewed~\cite{Gerhold:2007yb,Gerhold:2007gx,Fodor:2007fn,Gerhold:2007pj}.

Before, however, questions of the Higgs boson mass bounds and decay properties can be addressed, 
the phase structure of the model needs to be investigated in order to determine 
the (bare) coupling constants in parameter space where eventual calculations of 
phenomenological interest can be performed. Much effort has therefore been spent on 
phase structure investigations of lattice Higgs-Yukawa models in the past, see e.g. 
Refs.~\cite{Hasenfratz:1989jr,Lee:1989mi,Bock:1990tv,Lin:1991cs,Hasenfratz:1991it,Hasenfratz:1992xs,Bock:1992yr,Bock:1997fu,Bock:1999qa} 
for an only partial list. The phase structure of the new, chirally invariant Higgs-Yukawa
model has been studied analytically by means of a large $N_f$ computation~\cite{Gerhold:2007yb, Fodor:2007fn}, 
where $N_f$ denotes the number of fermion generations, as well as numerically by direct
Monte-Carlo simulations~\cite{Gerhold:2007gx, Fodor:2007fn}. 

In the present paper we study the dependence of the Higgs boson mass on the model parameters by direct Monte-Carlo simulations.
We confirm that the smallest and largest Higgs boson masses are obtained at vanishing bare
quartic self-coupling and at infinite bare quartic coupling, respectively, as expected
from perturbation theory. We then present our results on the cutoff-dependence of the lower Higgs boson mass 
bound and examine finite volume effects. The resulting finite volume Higgs boson mass bounds 
are extrapolated to the infinite volume limit. Since the aforementioned results
were determined in the mass degenerate case, \ie with equal top and bottom quark masses,
we also investigate the effect of the top-bottom mass splitting on the Higgs boson mass.
All these numerical findings are compared to corresponding
analytical calculations based on the effective potential at the 1-loop order of lattice perturbation theory.
The very good observed agreement between these two approaches justifies deriving the lower mass 
bound analytically from the effective potential also in the numerically very demanding, physical situation, 
where the bottom quark is approximately 40 times lighter 
than the top quark. Finally, we give a brief outlook towards the upper Higgs boson mass bound.

\section{The $\mbox{SU}(2)_L\times \mbox{U}(1)_R$ lattice Higgs-Yukawa model}
\label{sec:model}

The model we consider here, is a four-dimensional, chirally invariant 
$SU(2)_L \times U(1)_R$ lattice Higgs-Yukawa model~\cite{Luscher:1998pq}, aiming at the 
implementation of the chiral Higgs-fermion coupling structure of the pure Higgs-Yukawa 
sector of the Standard Model reading
\beq
\label{eq:StandardModelYuakwaCouplingStructure}
L_Y = y_b \left(\bar t, \bar b \right)_L \varphi b_R 
+y_t \left(\bar t, \bar b \right)_L \tilde\varphi t_R  + c.c.,
\eeq
with $y_{t,b}$ denoting the bare top and bottom Yukawa coupling constants.
Here, we have restricted ourselves to the consideration of the top-bottom
doublet $(t,b)$ interacting with the complex Higgs doublet $\varphi$ ($\tilde \varphi = i\tau_2\varphi^*$, and $\tau_i$ are the Pauli-matrices), 
since the Higgs dynamics is dominated by the coupling to the heaviest fermions. 
For the same reason we also {\it neglect all gauge fields} in this approach. 

The fields considered in this model are one four-component, real Higgs field $\Phi$, being equivalent to the
complex doublet $\varphi$ of the Standard Model, and $N_f$ top-bottom
doublets represented by eight-component spinors $\psi^{(i)}\equiv (t^{(i)}, b^{(i)})$, $i=1,...,N_f$.

The chiral character of the targeted coupling structure in Eq.~(\ref{eq:StandardModelYuakwaCouplingStructure}) 
can be preserved on the lattice by constructing the fermionic action $S_F$ from the (doublet) Neuberger overlap 
operator $\D$~\cite{Neuberger:1998wv} according to
\bea
S_F &=& \sumFL\,
\bar\psi^{(i)}\, \fermiMat\, \psi^{(i)}, \\
\label{eq:DefYukawaCouplingTerm}
\fermiMat &=& \D + 
P_+ \phi^\dagger \fhs{1mm}\mbox{diag}\left(\hat y_t,\hat y_b\right) \hat P_+
+ P_- \fhs{1mm}\mbox{diag}\left(\hat y_t,\hat y_b\right) \phi \hat P_-.
\eea
Here the Higgs field $\Phi_x$, defined on the site indices $x=(\vec x, t)$ of the $L_s^3\times L_t$-lattice,
was written as a quaternionic, $2 \times 2$ matrix 
$\phi_x = \Phi_x^\mu \sigma_\mu,\, \sigma_0=\ID,\, \sigma_j =  -i\tau_j$  with $\vec\tau$ denoting the vector 
of Pauli matrices, acting on the flavour index of the fermionic doublets.
The left- and right-handed projection operators $P_{\pm}$ and the modified projectors $\hat P_{\pm}$
are given as
\bea
P_\pm = \frac{1 \pm \gamma_5}{2}, \quad &
\hat P_\pm = \frac{1 \pm \hat \gamma_5}{2}, \quad &
\hat\gamma_5 = \gamma_5 \left(\ID - \frac{1}{\rho} \D \right),
\eea
with $\rho$ being the radius of the circle of eigenvalues in the complex plane of the free
Neuberger overlap operator~\cite{Neuberger:1998wv}.
This action now obeys an exact $\mbox{SU}(2)_L\times \mbox{U}(1)_R$ lattice chiral symmetry. For $\Omega_L\in \mbox{SU}(2)$
and $U_R \in U(1)$ the action is invariant under the transformation
\bea
\label{eq:ChiralSymmetryTrafo1}
\psi\rightarrow  U_R \hat P_+ \psi + \Omega_L \hat P_- \psi,
&\quad&
\bar\psi\rightarrow  \bar\psi P_+ \Omega_L^\dagger + \bar\psi P_- U^\dagger_R, \\
\label{eq:ChiralSymmetryTrafo2}
\phi \rightarrow  U_R  \phi \Omega_L^\dagger,
&\quad&
\phi^\dagger \rightarrow \Omega_L \phi^\dagger U_R^\dagger.
\eea
Note that in the mass-degenerate case, \ie $\hat y_t=\hat y_b$, this symmetry is extended to $\mbox{SU}(2)_L\times \mbox{SU}(2)_R$.
In the continuum limit the symmetry given in Eq.~(\ref{eq:ChiralSymmetryTrafo1}) and Eq.~(\ref{eq:ChiralSymmetryTrafo2}) recovers the continuum 
$\mbox{SU}(2)_L\times \mbox{U}_R(1)$ chiral symmetry and the lattice Higgs-Yukawa coupling becomes 
equivalent to the coupling structure in Eq.~(\ref{eq:StandardModelYuakwaCouplingStructure}) when identifying 
\bea
\varphi_x = 
C\cdot \left(
\begin{array}{*{1}{c}}
\Phi_x^2 + i\Phi_x^1\\
\Phi_x^0-i\Phi_x^3\\
\end{array}
\right),\quad
&
\tilde\varphi_x = i\tau_2\varphi^*_x = 
C\cdot \left(
\begin{array}{*{1}{c}}
\Phi_x^0 + i\Phi_x^3\\
-\Phi_x^2+i\Phi_x^1\\
\end{array}
\right), \quad
&
\mbox{and} \quad
y_{t,b} = \frac{\hat y_{t,b}}{C} \quad
\eea
for some real, non-zero constant $C$. Note also that in absence of gauge fields the Neuberger Dirac operator can be trivially 
constructed in momentum space, since its eigenvectors and eigenvalues $\nu^\epsilon(p),\, p\in\ImpSpace$ are explicitly known. 
Here, $\ImpSpace$ is the set of all lattice momenta and $\epsilon=\pm 1$ is the sign of the imaginary part of the eigenvalues
coming in complex conjugate pairs. In this notation they are given as~\cite{Gerhold:2007yb}
\bea
\label{eq:eigenValOfFreeND}
\nu^\epsilon(p)&=& \rho + \rho\cdot\frac{\epsilon i\sqrt{\tilde p^2} + \frac{r}{2}\hat p^2 - \rho}{\sqrt{\tilde p^2 + (\frac{r}{2}\hat p^2 -
\rho)^2}},\quad \tilde p^2 = \sum\limits_{\mu=0}^3 \sin^2(p_\mu),\quad 
\hat p^2 = 4\sum\limits_{\mu=0}^3 \sin^2\left(\frac{p_\mu}{2}\right)
\eea
where $r$ denotes the coefficient of the Wilson term in the underlying Wilson Dirac operator.

Finally, the lattice Higgs action $S_{\Phi}$ is given by the lattice $\Phi^4$-action
\beq
\label{eq:LatticePhiAction}
S_\Phi = -\hat\kappa\sum_{x,\mu} \Phi_x^{\dagger} \left[\Phi_{x+\hat\mu} + \Phi_{x-\hat\mu}\right]
+ \sum_{x} \Phi^{\dagger}_x\Phi_x + \hat\lambda \sum_{x} \left(\Phi^{\dagger}_x\Phi_x - N_f \right)^2,
\eeq
which is equivalent to the lattice action in continuum notation
\bea
\label{eq:ContinuumPhiAction}
S_\varphi &=& \sum_{x} \left\{ \frac{1}{2} \left(\nabla^f_\mu\varphi\right)_x^{\dagger} \nabla^f_\mu\varphi_x
+ \frac{1}{2} m^2\varphi_x^{\dagger}\varphi_x + \lambda \left(\varphi_x^{\dagger}\varphi_x\right)^2   \right\},
\eea
with the bare mass $m$, the bare quartic self-coupling constant $\lambda$, and $\nabla^f_\mu$ denoting
the lattice forward derivative in direction $\mu$. The connection is
established through a rescaling of the Higgs field and the involved coupling constants according
to
\beq
\label{eq:RelationBetweenHiggsActions}
\varphi_x = \sqrt{2\hat\kappa}
\left(
\begin{array}{*{1}{c}}
\Phi_x^2 + i\Phi_x^1\\
\Phi_x^0-i\Phi_x^3\\
\end{array}
\right),
\quad
\lambda=\frac{\hat\lambda}{4\hat\kappa^2}, \quad
m^2 = \frac{1 - 2N_f\hat\lambda-8\hat\kappa}{\hat\kappa}, \quad
y_{t,b} = \frac{\hat y_{t,b}}{\sqrt{2\hat\kappa}}.
\eeq

In the given model the expectation value $\langle \varphi \rangle$ would always be identical to zero due to the 
symmetries given in Eq.~(\ref{eq:ChiralSymmetryTrafo1}) and Eq.~(\ref{eq:ChiralSymmetryTrafo2}). To study 
the mechanism of spontaneous symmetry breaking, one usually introduces an external current, which has 
to be removed after taking the thermodynamic limit, leading then to the existence of symmetric and broken 
phases with respect to the order parameter $\langle \varphi\rangle$ as desired. An alternative approach, which was 
shown to be equivalent in the thermodynamic limit~\cite{Hasenfratz:1989ux,Hasenfratz:1990fu,Gockeler:1991ty}, 
is to use the global symmetries of the model to rotate each field configuration $(\varphi_x)$ after its 
generation in the Monte-Carlo process according to
\beq
\label{eq:GaugeRotation}
\varphi^{rot}_x = U \varphi_x
\eeq
with the $\mbox{SU}(2)$ matrix $U$ selected for each field configuration $(\varphi_x)$
such that 
\beq
\label{eq:GaugeRotationRequirement}
\sum\limits_x \varphi_x^{rot} = 
\left(
\begin{array}{*{1}{c}}
0\\
\left|\sum\limits_x \varphi_x \right|\\
\end{array}
\right).
\eeq
Here, we use this second approach. According to the notation used in Eq.~(\ref{eq:ContinuumPhiAction}), which already 
incorporates a factor $1/2$ in contrast to the standard notation, the relation between 
the bare vacuum expectation value of the Higgs mode $v$ and the expectation value of $\varphi^{rot}$
is then given as 
\beq
\label{eq:DefOfVEV}
\langle \varphi^{rot} \rangle = 
\left(
\begin{array}{*{1}{c}}
0\\
v\\
\end{array}
\right).
\eeq
In this approach the unrenormalized Higgs field $h_x$ and the Goldstone fields $g^1_x,g^2_x,g^3_x$ can 
directly be read out of the rotated field $\varphi_x^{rot}$ according to
\beq
\label{eq:DefOfHiggsAndGoldstoneModes}
\varphi_x^{rot}  = 
\left(
\begin{array}{*{1}{c}}
g_x^1 + ig_x^2\\
v + h_x - i g_x^3\\
\end{array}
\right).
\eeq

\section{Simulation strategy and observables}

The general method we apply to determine the lower and upper Higgs boson mass bounds
is the numerical evaluation of the whole range of Higgs boson masses that are attainable
within our model at a fixed value of the cutoff in consistency with phenomenology. The latter requirement
restricts the freedom in the choice of the bare parameters $\hat \kappa, \hat y_{t,b}, \hat \lambda$ 
due to the phenomenological knowledge of the renormalized vacuum expectation value of
the Higgs mode (vev), \ie $v_r/a = 246\, \mbox{GeV}$, and of the top and bottom quark masses, 
\ie $m_{t}/a \approx 175\, \mbox{GeV}$ and $m_{b}/a \approx 4.2\, \mbox{GeV}$,
respectively. Here, $m_t$, $m_b$ denote the lattice masses and $a$ is the lattice spacing.
For our numerical simulations we use the tree-level relation 
\beq
\label{eq:treeLevelTopMass}
y_{t,b} = \frac{m_{t,b}}{v_r}
\eeq
as a first guess to set the bare Yukawa coupling constants $y_t$ and $y_b$. The actually 
resulting quark masses are discussed in section~\ref{sec:LowerMassBound}.
Furthermore, the model has to be evaluated in the broken phase, \ie at non-vanishing
vacuum expectation value of the Higgs mode, $v \neq 0$, however close to the phase transition 
to the symmetric phase. We also use the phenomenologically known value of the renormalized 
vev to determine the lattice spacing $a$ and thus the physical cutoff $\Lambda$ 
according to
\bea
246\, \mbox{GeV} = \frac{v_r}{a} \equiv \frac{v}{\sqrt{Z_G}\cdot a}, \quad&
\Lambda = a^{-1}, \quad&
\left[\tilde G_G(\hat p^2)\right]^{-1} = \frac{\hat p^2+m_{Gp}^2}{Z_G}.
\eea
Here, the Goldstone propagator mass $m_{Gp}$ is zero except for finite volume contributions and
the renormalization constant $Z_G$ is obtained from the Goldstone propagator $\tilde G_G(\hat p^2)$ 
in momentum space with $\hat p^2$ denoting the squared lattice momenta. The Higgs propagator mass $m_{Hp}$
and the corresponding renormalization factor $Z_H$ are obtained in the same way through the relation
\beq
\label{eq:DefOfHiggsPropMass}
\left[\tilde G_H(\hat p^2)\right]^{-1} = \frac{\hat p^2+m_{Hp}^2}{Z_H}
\eeq
where the Goldstone and Higgs propagators in position space are given as
\bea
\label{eq:DefOfPropsInPosSpace}
G_G(x,y) = \frac{1}{3}\sum\limits_{\alpha=1}^3 \langle g^\alpha_x g^\alpha_y \rangle 
&\quad \mbox{and} \quad&
G_H(x,y) = \langle h_x h_y \rangle.
\eea

The aforementioned Higgs propagator mass is connected via perturbation theory to the actual physical Higgs 
boson mass $m_H$ given by the corresponding energy eigenvalue of the Hamiltonian. It can
be obtained from the exponential decay of the Higgs time slice correlation function
\bea
\label{eq:DefOfHiggsTimeSliceCorr}
C_H(\Delta t) = \frac{1}{L_t\cdot L_s^6} \sum\limits_{t=1}^{L_t} \sum\limits_{\vec x, \vec y}
\langle h_{\vec x,t}\cdot h_{\vec y, t+\Delta t} \rangle
&\,\mbox{with}\,&
C_H(L_t/2>\Delta t \gg 1) \propto \mbox{cosh}\Big[m_H\cdot (L_t/2-\Delta t)\Big].\quad\quad
\eea
Correspondingly, we derive the physical top and bottom quark masses $m_t, m_b$ from the fermionic time slice correlators
\bea
\label{eq:DefOfFermionTimeSliceCorr}
C_f(\Delta t) &=& \frac{1}{L_t\cdot L_s^6} \sum\limits_{t=1}^{L_t} \sum\limits_{\vec x, \vec y}
\Big\langle \mbox{Tr}\,\left(f_{\vec x, t}\cdot \bar f_{\vec y, t+\Delta t}\right) \Big\rangle
\eea
with $f=t,b$ selecting the quark flavour.
We remark here that the full {\it all-to-all} correlator as defined in Eq.~(\ref{eq:DefOfFermionTimeSliceCorr}) can be 
trivially computed by using sources which have a constant value of one on a whole time slice for a selected
spinor index and a value of zero everywhere else.
This all-to-all correlator yields very clean signals for the top and bottom quark mass determination.

For a given cutoff $\Lambda$ the aforementioned requirements of reproducing the vev and the quark masses still leave open 
a one-dimensional freedom in the model parameters, which can be parametrized in terms of the quartic self-coupling constant $\lambda$. 
However, aiming at lower and upper Higgs boson mass bounds,
this remaining freedom can be fixed, since it is expected from perturbation theory that the lightest
Higgs boson masses are obtained at vanishing self-coupling constant $\lambda=0$, and the heaviest masses at infinite coupling constant $\lambda=\infty$,
respectively, according to the qualitative one-loop perturbation theory result for the Higgs boson mass shift
\beq
\label{eq:perturbTheroyResult}
\delta m_H^2/a^2 \equiv (m_H^2 - m^2)/a^2 \propto  \left(\lambda - y_t^2 - y_b^2 \right) \cdot \Lambda^2.
\eeq
Subsequently, it will be explicitly checked by direct lattice calculations that the lowest and highest Higgs boson masses are indeed obtained
at vanishing  and infinite bare quartic coupling constants, respectively.
The lower Higgs boson mass bound will then be searched for in the weak self-coupling region, \ie at $\lambda\ll 1$,
while the upper bound will be determined from simulations performed in the strong coupling region, \ie at $\lambda\gg 1$.

For the numerical evaluation of the model we have implemented a PHMC-algorithm~\cite{Frezzotti:1997ym,Frezzotti:1998eu,Frezzotti:1998yp}, 
allowing access to the physical situation of odd $N_f$. We make heavy use of the fact
that no gauge fields are considered here. In this setup the Dirac operator is
diagonal in momentum space and its eigenvalues and eigenvectors are explicitly known.
In our approach we use a fast Fourier transformation~\cite{FFTW05} to switch between momentum space, where
$\D$ can be trivially applied, and position space, where we perform the multiplication with the field $\Phi_x$.
Besides that, it was crucial to reduce the auto-correlation times 
of the measured observables, which we achieved with the help of Fourier acceleration~\cite{Batrouni:1985jn, book:ParisiFACC}.
We also remark here that the condition number of the fermionic matrix could greatly be reduced
by preconditioning. However, the details of the algorithm will be discussed elsewhere.

All results in the following have been obtained at $N_f=1$ with degenerate Yukawa coupling constants, 
\ie $y_t=y_b$ (unless otherwise stated), owing to the much higher numerical requirements at the realistic
ratio $y_b/y_t=0.024$ exceeding our available resources.
However, on some small lattice sizes we could also investigate the dependence of
the obtained Higgs boson masses on the top-bottom mass splitting ratio $y_b/y_t$, allowing ultimately to extrapolate
our results to the physical scenario. Additionally, we are currently working on the $N_f=3$ results to account 
for the colour index (even though gauge fields are still absent).

\section{$\lambda$-dependence of the Higgs boson mass and effective potential}

We now turn to the determination of the cutoff dependent lower Higgs boson mass bound. As discussed in the
previous section, one would expect the lightest Higgs boson masses at vanishing quartic self-coupling parameter $\lambda=0$.
However, one should remark here that the given argument is not complete, since the phase transition line changes
while varying $\lambda$ as well, thus making the bare Higgs boson mass $m$ a function of $\lambda$ for fixed cutoff $\Lambda$
and fixed Yukawa coupling constants.
In fact, the phase transition line is strongly shifted when varying the self-coupling parameter~\cite{Gerhold:2007yb}.
From Eq.~(\ref{eq:perturbTheroyResult}) alone one can therefore not conclude whether the lightest Higgs boson mass in this model
will actually be observed at $\lambda=0$ or rather at slightly stronger self-coupling.

To examine this we consider the effective potential~\cite{Coleman:1973jx} in terms
of the vacuum expectation value $v$. In the large $N_f$-limit with $\hat y_{t,b}= O(1/\sqrt{N_f})$
and $\hat \lambda = O(1/N_f)$ it was given for this model in Ref.~\cite{Gerhold:2007yb} as
\bea
\label{eq:EffPot}
U[v] &=&
\frac{1}{2} m^2 v^2 + \lambda v^4 + U_{F}[v] \\
U_{F}[v] &=&
\frac{-2N_f}{L_s^3\cdot L_t}\cdot \sum\limits_{p\in\ImpSpace} \log\left|\nu^+(p) + y_t v \left(1-\frac{1}{2\rho}\nu^+(p)\right)  \right|^2
+ \log\left|\nu^+(p) + y_b v \left(1-\frac{1}{2\rho}\nu^+(p)\right)  \right|^2. \quad\quad
\label{eq:FermionEffectivePot}
\eea
For a given value of $v$, and thus for a fixed cutoff $\Lambda$, the dependence of the bare Higgs boson mass on the
quartic self-coupling constant can be trivially determined by setting the first derivative of $U[v]$ to zero according to
\beq
\label{eq:BareMassFromU}
m^2 = -4\lambda v^2 - \frac{1}{v} \frac{\mbox{d}}{\mbox{d}v} U_F[v].
\eeq
Correspondingly, one can also estimate the Higgs propagator mass from the curvature of the effective potential yielding
\beq
\label{eq:PropMassFromU}
m^2_{Hp} = 8 \lambda v^2 - \frac{1}{v} \frac{\mbox{d}}{\mbox{d}v} U_F[v] + \frac{\mbox{d}^2}{\mbox{d}v^2} U_F[v].
\eeq
From this result it is clear that the lightest Higgs boson masses should be observed at vanishing quartic coupling constant, 
since the fermionic contribution $U_F$ does not depend on $\lambda$ at the considered order.
This observation allows to restrict the search for the lower Higgs boson mass bounds to the 
setting $\lambda = 0$ in the following. 

\begin{figure}[htb]
\begin{tabular}{cc}
\includegraphics[width=0.48\textwidth]{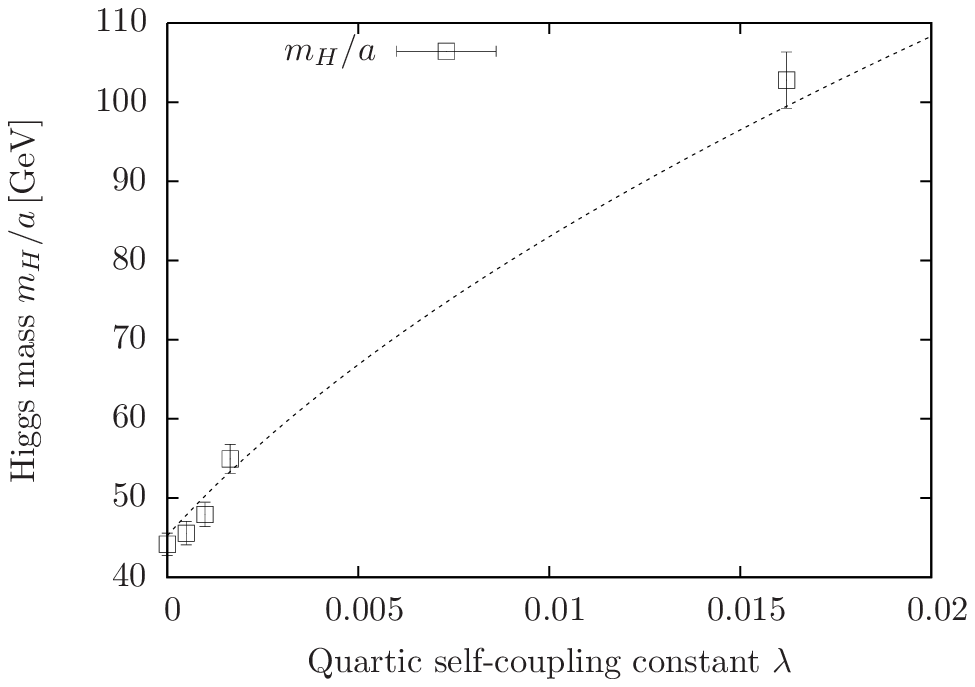}
&
\includegraphics[width=0.48\textwidth]{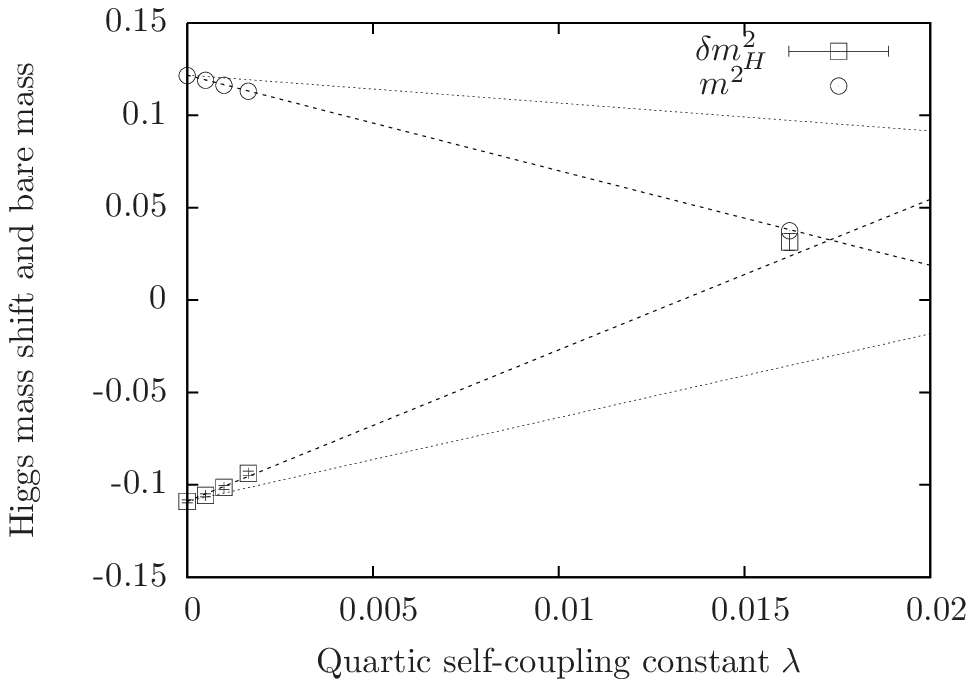}\\
\hs{4mm}(a) & \hs{8mm}(b)  \\
\end{tabular}
\caption{(a) The dependence of the Higgs boson mass $m_H/a$ on the quartic self-coupling constant $\lambda$ at 
$\Lambda= 400\, \mbox{GeV}$ on a $16^3\times 32$-lattice for constant, degenerate Yukawa coupling constants
fixed according to Eq.~(\ref{eq:treeLevelTopMass}) and $N_f=1$. 
The prediction from the effective potential in 
Eq.~(\ref{eq:EffPot}) is represented by the dashed curve.
(b) The corresponding Higgs boson mass shifts $\delta m_H^2$ and bare masses $m^2$ versus the quartic coupling 
constant $\lambda$. The dotted (dashed) lines show the analytical results derived from the unimproved (improved) 
effective potential.}
\label{fig:lambdaDependence}
\vs{-2mm}
\end{figure}

We check the latter result by studying the $\lambda$-dependence of the Higgs boson mass in direct Monte-Carlo simulations performed on 
a $16^3\times 32$-lattice for constant cutoff $\Lambda=400\, \mbox{GeV}$, $N_f=1$, and degenerate Yukawa coupling parameters fixed according to
Eq.~(\ref{eq:treeLevelTopMass}). In Fig.~\ref{fig:lambdaDependence}a we show the Higgs 
correlator mass $m_H/a$, which is almost identical to the propagator mass in the weak coupling region, versus the quartic coupling 
constant $\lambda$. The mass monotonously falls with $\lambda$ going to zero and converges to the Higgs boson mass obtained at 
$\lambda=0$. We remark that at $\lambda=0$ the model is still well defined even in the broken phase, at least sufficiently close to the 
phase transition line thanks to the squared bare mass $m^2$ being positive provided that the Yukawa coupling constants are non-zero.
Furthermore, the numerical results are in good agreement with the analytical predictions in Eq.~(\ref{eq:PropMassFromU}),
where the renormalization constant $Z_G$ has been set to one for the conversion between $\Lambda$ and $v$. 

In Fig.~\ref{fig:lambdaDependence}b we additionally present the numerical results for the mass shifts $\delta m_H^2$ together with the corresponding
squared bare masses $m^2$. One sees that 
the bare mass becomes smaller with increasing self-coupling constant $\lambda$, but that this effect is overcompensated
by the much stronger rise of the mass shift $\delta m_H^2$, as expected from the predictions in Eq.~(\ref{eq:BareMassFromU}) and Eq.~(\ref{eq:PropMassFromU})
marked by the dotted lines. However, on a quantitative level there is a huge discrepancy between the analytical predictions and the numerical
results concerning both, the mass shifts as well as the bare masses. 

This discrepancy can be fixed by taking the next order in the loop expansion of the effective potential~\cite{Coleman:1973jx} of the 
lattice Higgs action in Eq.~(\ref{eq:ContinuumPhiAction}) into account, which would correspond to the $1/N_f$-correction  of the pure
bosonic part of the effective potential in the language of the large $N_f$ analysis. This next to leading order contribution is given by connecting two 
legs of the four-point vertex and summing up the arising scalar propagator over all lattice momenta yielding the improved effective potential
\beq
\label{eq:ImprovedEffPot}
U_{imp}[v] = U[v] + \lambda v^2 \frac{1}{L_s^3\cdot L_t}\left[\sum\limits_{p\in\ImpSpace} \frac{6}{\hat p^2+m_{Hp}^2} 
+ \sum\limits_{0\neq p\in\ImpSpace} \frac{6}{\hat p^2+m_{Gp}^2} \right] .
\eeq
\bc
\ec
Here, the momentum $p=0$ is excluded in the sum over the Goldstone
loop, since the Goldstone propagator at zero momentum is identical to zero due to the rotation of the field $\varphi$ introduced in 
section~\ref{sec:model}. However, with increasing lattice volume this exclusion 
looses any significance. The factors appearing in the nominators of Eq.~(\ref{eq:ImprovedEffPot}) are given by the multiplicities of the
considered diagrams. Though the effective potential $U[v]$ was actually given in bare constants, we plug the renormalized Higgs
and Goldstone boson masses into the expressions for the scalar propagators, in order to further improve the result. 
For the evaluation of Eq.~(\ref{eq:ImprovedEffPot}) independent of any knowledge from Monte-Carlo simulations we use $m_{Gp} = 0$  in the following
and determine $m_{Hp}$ self-consistently with respect to the curvature of the resulting improved effective potential at its minimum.

The results for the mass shifts and bare masses obtained by the improved effective potential are shown in Fig.~\ref{fig:lambdaDependence}b
depicted by the dashed lines. The improved results are in very good agreement with the numerical data. The reason why the unimproved effective
potential could describe the renormalized masses presented in Fig.~\ref{fig:lambdaDependence}a so well - in contrast to the bare masses and mass shifts - is,
that the improvement term exactly cancels in Eq.~(\ref{eq:PropMassFromU}) but not in the equation for the bare mass.

\section{Lower mass bounds}
\label{sec:LowerMassBound}

Given the knowledge about the $\lambda$-dependence of the renormalized Higgs boson mass we can now safely determine the cutoff-dependence of the lower Higgs 
boson mass bound by evaluating the Higgs boson mass at $\lambda=0$ for several values of $\Lambda$. Two restrictions limit the
range of accessible energy scales: on the one side all particle masses have to be small 
compared to $\Lambda$ to avoid unacceptably large cutoff-effects, on the other side all masses have to be large
compared to the inverse lattice size to avoid finite volume effects. As a minimal requirement 
we demand here that all particle masses $\hat m$ in lattice units fulfill 
\bea
\label{eq:RequirementsForLatMass}
\hat m < 0.5& \quad \mbox{and} \quad & \hat m\cdot L_{s,t}>2. 
\eea
For a lattice with side lengths $L_s=L_t=32$, a degenerate top/bottom quark mass of $175\, \mbox{GeV}$, 
and Higgs boson masses ranging from $40$ to $70$ GeV one can thus access energy scales $\Lambda$ from $350\,\mbox{GeV}$ 
to approximately $1100$ GeV. 

\begin{figure}[htb]
\begin{tabular}{cc}
\includegraphics[width=0.557\textwidth]{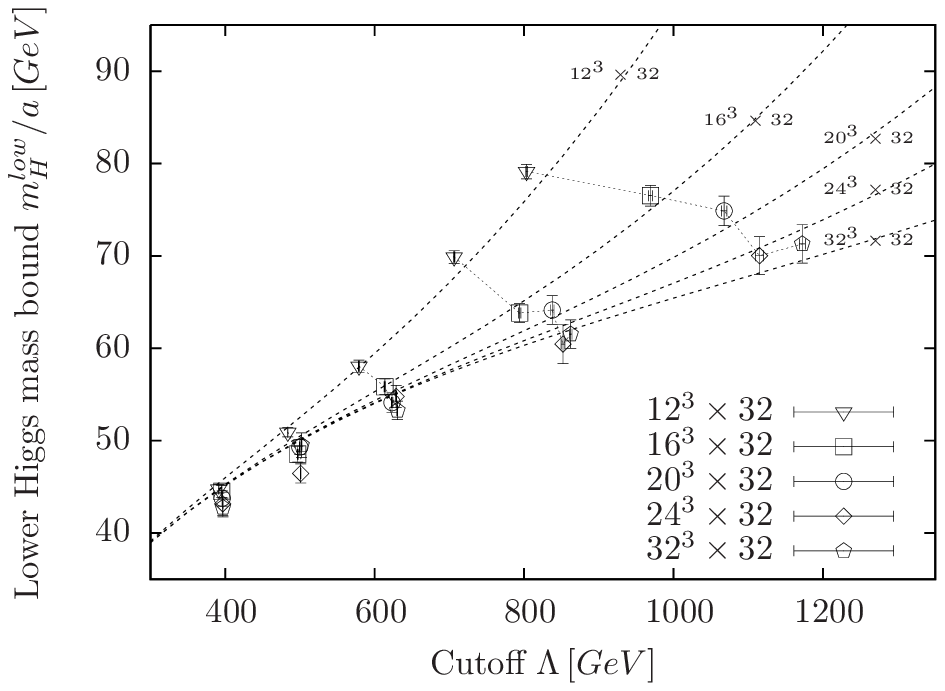} 
&
\includegraphics[width=0.443\textwidth]{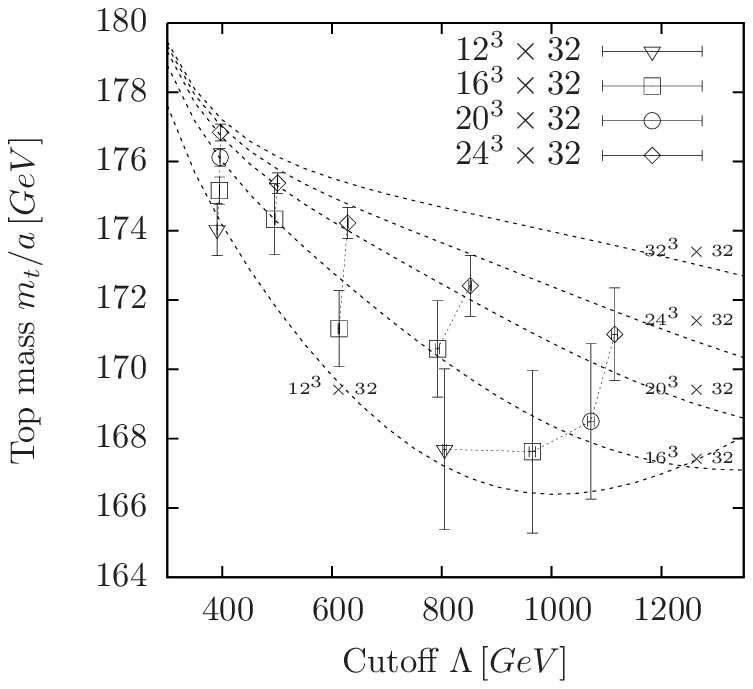} \\
\hs{4mm}(a) & \hs{8mm}(b)  \\
\end{tabular}
\caption{(a) The lower Higgs boson mass bound $m_H^{low}/a$ versus the cutoff $\Lambda$.
(b) The cutoff dependence of the top quark mass $m_t/a$.
To illustrate finite volume effects, simulations have been rerun with identical parameter sets but different
lattice sizes. Runs with same parameter sets are connected via dotted lines to guide the eye.
The analytical predictions for the respective lattice sizes derived from the effective potential are depicted by the dashed curves.
All presented results have been obtained in Monte-Carlo simulations with identical, degenerate
bare Yukawa coupling constants fixed according to Eq.~(\ref{eq:treeLevelTopMass}), $\lambda=0$, and $N_f=1$.}
\label{fig:LowerBound}
\vs{-2mm}
\end{figure}

In Fig.~\ref{fig:LowerBound}a we show the numerically obtained Higgs correlator masses $m_H/a$ versus the cutoff $\Lambda$. 
All presented results have been obtained in Monte-Carlo simulations with identical, degenerate
bare Yukawa coupling constants fixed according to Eq.~(\ref{eq:treeLevelTopMass}), $\lambda=0$, and $N_f=1$.
To illustrate the influence of the finite lattice volume we have rerun all simulations with exactly the
same parameter settings but different lattice sizes. Those results belonging to the same parameter
sets are connected by dotted lines to guide the eye. Additionally, we also present the corresponding analytical, finite volume 
expectations for the cutoff-dependence of the Higgs boson mass as derived from the effective potential. 
We observe that the effective potential calculation is in very good agreement with the cutoff dependence of the measured Higgs boson masses.
It also correctly predicts the observed finite volume effects. 

The cutoff dependence of the corresponding top quark masses is presented in Fig.~\ref{fig:LowerBound}b. Those results belonging
to the same bare parameters are connected by dotted lines for a better overview. The renormalized top quark mass, and thus the renormalized Yukawa
coupling constant, decreases slightly with growing cutoff as expected in a trivial theory. In order to hold the fermion masses constant as
desired one would have to fine-tune the bare Yukawa coupling parameters, giving rise to an additional but small contribution to the 
cutoff dependence of the Higgs boson mass. 

For an efficient tuning of the bare Yukawa coupling constants, however, a thorough analytical understanding of the observed behaviour
of the fermion mass would be indispensable. We therefore compare the presented numerical findings for the top quark mass
to the corresponding 1-loop lattice perturbation theory predictions, depicted by the dashed lines in Fig.~\ref{fig:LowerBound}b
and observe good agreement, allowing one ultimately to hold the fermion masses constant in follow-up Monte-Carlo 
calculations.

\bc
\ec

The presented perturbative results are given by the tree-level fermion propagator plus the 1-Goldstone-loop and 1-Higgs-loop corrections according to
\bea
\left(\begin{array}{*{2}{c}}
t \bar t(p)  &  \\
 & b \bar b(p)\\
\end{array}\right)
&=& \left[\tilde D(p) - \Sigma(p)  \right]^{-1}, \quad \tilde D(p) = \D(p) + v\cdot B_0 \Gamma(p), \\
\Gamma(p) &=& \ID - \frac{1}{2\rho} \D(p), \quad B_\mu = P_+ \sigma^\dagger_\mu \mbox{diag}(y_t, y_b) + P_-\mbox{diag}(y_t, y_b) \sigma_\mu,  \\
\Sigma(p) &=& \frac{1}{L_s^3\cdot L_t} \cdot  \sum\limits_{k\in\ImpSpace} \frac{B_0\Gamma(k)\tilde D^{-1}(k)B_0\Gamma(p)}
{\hat t^2 + m^2_{Hp}} 
\nonumber\\
&+ &
\frac{1}{L_s^3\cdot L_t} \cdot  \sum\limits_{p\neq k\in\ImpSpace}\sum\limits_{i=1}^3 
\frac{B_i\Gamma(k)\tilde D^{-1}(k)B_i\Gamma(p)}
{\hat t^2}, \quad \mbox{with}\quad t=p-k.
\eea
Here $\hat t^2$ is the squared lattice momentum of the relative momentum $t=p-k$ as defined in Eq.~(\ref{eq:eigenValOfFreeND}) and
$\D(p)$ denotes the $8 \times 8$ doublet Dirac matrix in momentum space for the given lattice momentum $p\in \ImpSpace$.
We remark that the contribution from the tadpole diagram to the fermion propagator is exactly canceled
by the renormalization of the vev induced by the tadpole diagram.
Again, the fact that the zero momentum mode of the Goldstone propagator is identical to zero due to the rotation of
the field $\varphi$ as described in section~\ref{sec:model} has to be respected by excluding these modes in the summation.
From the 1-loop propagators $t\bar t(p)$ and $b\bar b(p)$ we calculate the top and bottom time-slice correlator as defined in 
Eq.~(\ref{eq:DefOfFermionTimeSliceCorr}) via a Fourier transformation in the time direction. The physical fermion masses shown in
Fig.~\ref{fig:LowerBound}b are then determined from the exponential decay of this correlator. 

\begin{figure}[htb]
\begin{tabular}{cc}
\includegraphics[width=0.48\textwidth]{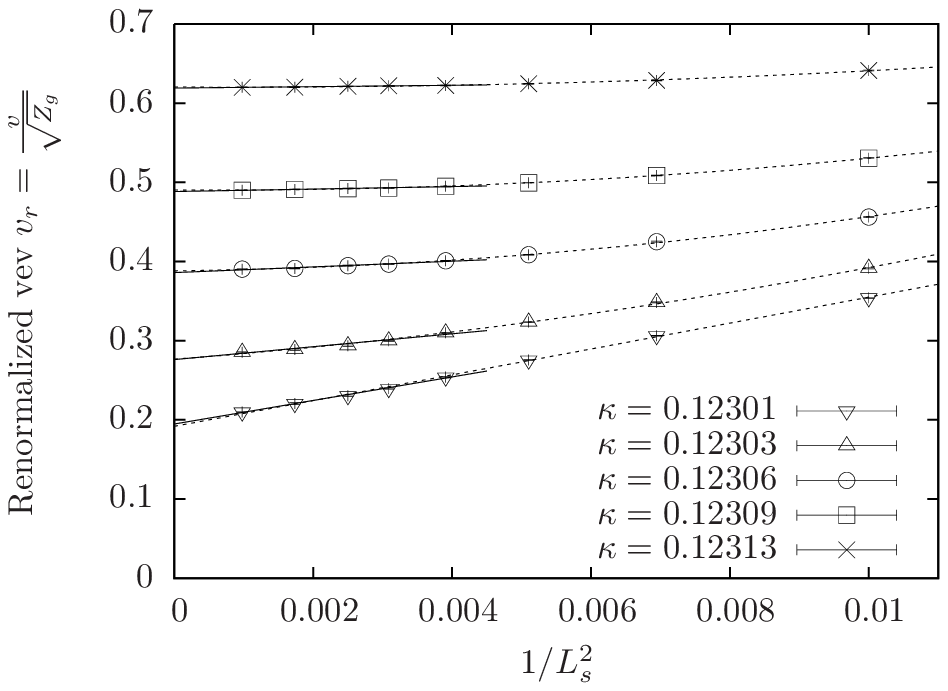}
&
\includegraphics[width=0.48\textwidth]{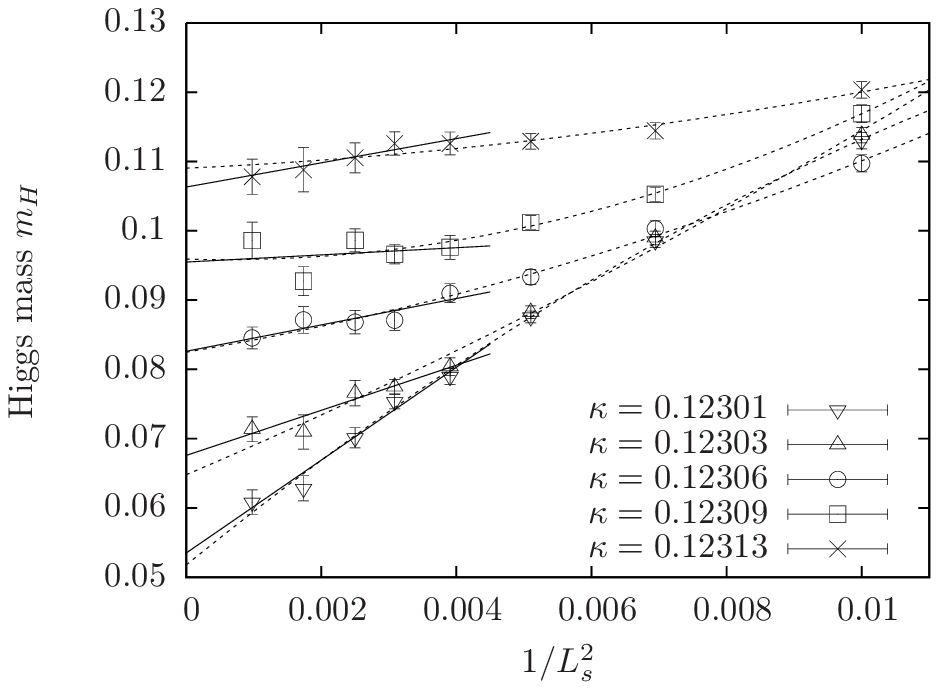}\\
\hs{4mm}(a) & \hs{8mm}(b)  \\
\end{tabular}
\caption{(a) The dependence of the renormalized vev $v_r = v/\sqrt{Z_G}$ on the squared inverse lattice side length $1/L^2_s$ for 
the same model parameters used in the Monte-Carlo runs presented in Fig.~\ref{fig:LowerBound}a but for a greater set of
different lattice sizes. The given hopping parameters $\kappa$ are 
associated to cutoffs ranging from $\Lambda=400\,\mbox{GeV}$ 
to $\Lambda\approx 1200\,\mbox{GeV}$. 
(b) The dependence of the Higgs correlator mass $m_H$ versus $1/L^2_s$ for 
the same Monte-Carlo runs.
The solid and dashed curves display the linear and parabolic fits as discussed in the main text. 
}
\label{fig:FiniteVolumeEffects}
\vs{-2mm}
\end{figure}

From Fig.~\ref{fig:LowerBound}a one also learns that the finite volume effects are rather mild at $\Lambda=400\, \mbox{GeV}$ 
with $m_H\cdot L_{s,t}>3.2$ on the $32^4$-lattice, while the vev, and thus the associated cutoff $\Lambda$, as well as the 
Higgs boson mass itself vary strongly with increasing lattice size $L_s$ at the higher cutoffs.
In Fig.~\ref{fig:FiniteVolumeEffects} we explicitly show this volume dependence of the vev and the Higgs boson mass, respectively. 

It is well known from investigations of pure $O(4)$ Higgs models~\cite{Hasenfratz:1989ux,Hasenfratz:1990fu,Gockeler:1991ty}
that the vev as well as the mass receive strong contributions from the massless Goldstone modes, inducing finite volume effects starting at 
$O(L_s^{-2})$. The next non-trivial finite volume contribution was shown to be of order $O(L_s^{-4})$. In Fig.~\ref{fig:FiniteVolumeEffects}
we therefore plot the obtained data versus $1/L_s^2$ and use the linear fit ansatz 
\beq
\label{eq:LinFit}
f^{(l)}_{v,m}(L_s^{-2}) = A^{(l)}_{v,m} + B^{(l)}_{v,m}\cdot L_s^{-2}
\eeq
to extrapolate to the infinite volume limit. The free fitting parameters $A^{(l)}_{v,m}$ and $B^{(l)}_{v,m}$
with the subscripts $v$ and $m$ refer to the renormalized vev $v_r$ and the Higgs boson mass $m_H$, respectively. 

\bc
\ec

To respect the presence of higher order terms in $1/L_{s}^{2}$ we include only the five largest lattice sizes in the linear fit. 
As a consistency check to test the dependence of these results on the choice of the fit ansatz, we furthermore consider the parabolic ansatz 
\beq
\label{eq:ParaFit}
f^{(p)}_{v,m}(L_s^{-2}) = A^{(p)}_{v,m} + B^{(p)}_{v,m}\cdot L_s^{-2} + C^{(p)}_{v,m}\cdot L_s^{-4}
\eeq
which we apply to the whole range of available lattice sizes. The deviation between the two fitting procedures are considered here as an 
additional systematic uncertainty for the infinite volume results determined from the linear fit in Eq.~(\ref{eq:LinFit}). 
The obtained fitting curves are displayed in Fig.~\ref{fig:FiniteVolumeEffects}a,b and the corresponding infinite volume extrapolations
of the vev and the Higgs boson mass are listed in Tab.~\ref{tab:ResultOfExtrapolation}.

\begin{table}[htb]
\centering
\begin{tabular}{|c|c|c|c|c|c|c|}
\hline
                 & \multicolumn{2}{c|}{Linear fit}           & \multicolumn{2}{c|}{Parabolic fit}      & \multicolumn{2}{c|}{Final result} \\ \hline
$\kappa$  	 & $A^{(l)}_v$         & $A^{(l)}_m$         & $A^{(p)}_v$        & $A^{(p)}_m$        &  $v_r$      & $m_H$  \\ \hline
$\,0.12301\,$  	 & $\, 0.1947(15)\, $  & $\, 0.0535(20)\, $  & $\, 0.1918(24)\, $ & $\, 0.0517(15)\, $ &  $\, 0.1947(15)(29)\, $  & $\, 0.0535(20)(18)\, $  \\ 
$\,0.12303\,$  	 & $\, 0.2761(24)\, $  & $\, 0.0676(13)\, $  & $\, 0.2766(21)\, $ & $\, 0.0648(23)\, $ &  $\, 0.2761(24)(05)\, $  & $\, 0.0676(13)(28)\, $  \\ 
$\,0.12306\,$  	 & $\, 0.3860(10)\, $  & $\, 0.0826(15)\, $  & $\, 0.3882(73)\, $ & $\, 0.0825(14)\, $ &  $\, 0.3860(10)(22)\, $  & $\, 0.0826(15)(01)\, $  \\ 
$\,0.12309\,$  	 & $\, 0.4883(03)\, $  & $\, 0.0955(35)\, $  & $\, 0.4903(03)\, $ & $\, 0.0959(24)\, $ &  $\, 0.4883(03)(20)\, $  & $\, 0.0955(35)(04)\, $  \\ 
$\,0.12313\,$  	 & $\, 0.6191(02)\, $  & $\, 0.1063(11)\, $  & $\, 0.6204(04)\, $ & $\, 0.1090(17)\, $ &  $\, 0.6191(02)(13)\, $  & $\, 0.1063(11)(27)\, $  \\ 
\hline
\end{tabular}
\caption{The results of the infinite volume extrapolation of the Monte-Carlo data for the renormalized vev $v_r$ and the Higgs boson mass $m_H$.
The extrapolation was performed by applying the linear fit ansatz in Eq.~(\ref{eq:LinFit}) and the parabolic ansatz in Eq.~(\ref{eq:ParaFit})
as described in the main text. The final results are taken from the linear fit. An additional, systematic uncertainty of the final results is specified
in the second pair of brackets taken from the deviation between the linear and parabolic fit results.
}
\label{tab:ResultOfExtrapolation}
\end{table}

\begin{figure}[htb]
\begin{tabular}{cc}
\includegraphics[width=0.48\textwidth]{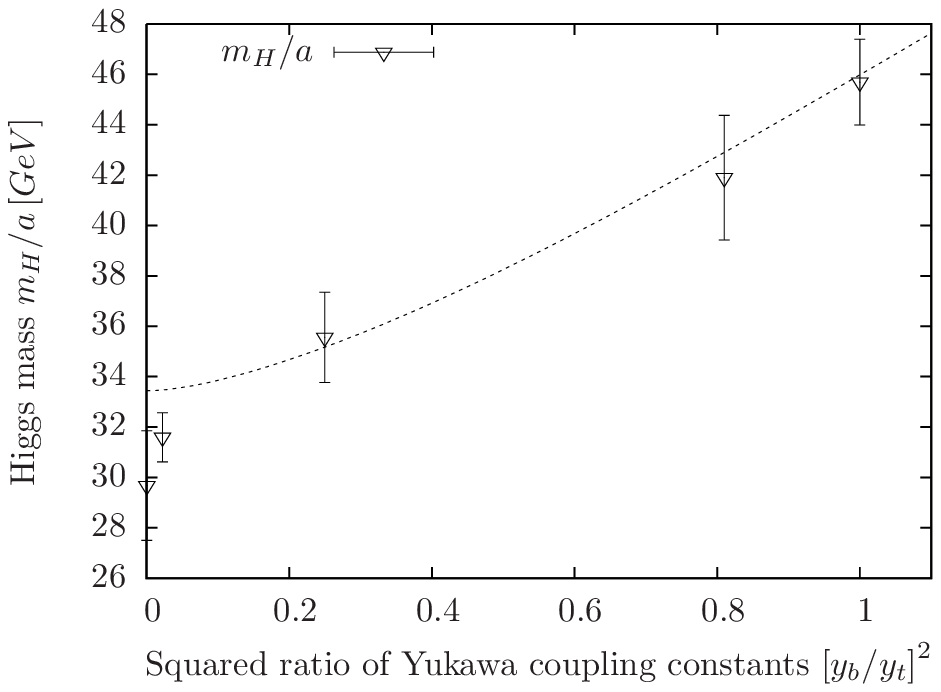}
&
\includegraphics[width=0.48\textwidth]{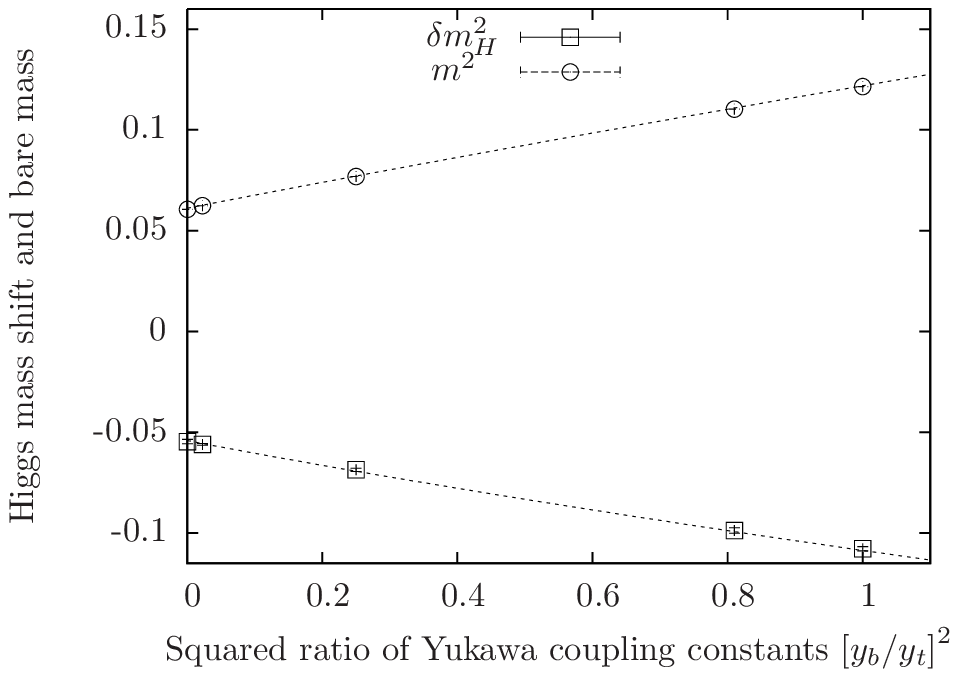}\\
\hs{4mm}(a) & \hs{8mm}(b)  \\
\end{tabular}
\caption{(a) The Higgs boson mass versus the squared ratio of the top and bottom Yukawa coupling constants 
$\left[y_b/y_t \right]^2$ on a $12^3\times 32$-lattice for constant cutoff $\Lambda=400\,\mbox{GeV}$,
$\lambda=0$, $N_f=1$, and $y_t$ fixed according to Eq.~(\ref{eq:treeLevelTopMass}). 
(b) The corresponding Higgs boson mass shifts $\delta m_H^2$ and bare masses $m^2$ versus $\left[y_b/y_t \right]^2$. 
The dashed curves represent the analytical predictions derived from the effective potential.}
\label{fig:SplitDependence}
\vs{-2mm}
\end{figure}

So far, the presented results have been determined in the mass degenerate case, \ie for $y_t=y_b$, which is 
easier to access numerically. This brings up the question of how the results are influenced when pushing the top-bottom
mass splitting to its physical value, \ie $m_b/m_t\approx 0.024$. From Eq.~(\ref{eq:perturbTheroyResult}) one
expects the Higgs boson mass shift $\delta m_H^2$ to grow quadratically with decreasing $y_b$ and that is
exactly what is observed in Fig.~\ref{fig:SplitDependence}b. Here, the bare top Yukawa coupling constant $y_t$, 
the quartic coupling parameter, and the cutoff are held constant, while $y_b$ is lowered to the physical 
ratio of $y_b/y_t$. 
The numerical data for the Higgs boson mass shift $\delta m_H^2$ are in excellent agreement with the predictions based on the
effective potential, which are depicted by the dashed lines in Fig.~\ref{fig:SplitDependence}b.
However, the Higgs boson mass itself does not increase but {\it decrease} with decreasing $y_b$ as shown in 
Fig.~\ref{fig:SplitDependence}a. This is because the first effect in the mass shift is over-compensated 
by the shift in the phase transition line, which is moved towards smaller bare Higgs boson masses $m$.
This behaviour of the bare Higgs boson mass is demonstrated also in Fig.~\ref{fig:SplitDependence}b, where the
numerical findings for the bare mass are compared to the corresponding expectations from the effective potential.
Again very good agreement between both approaches is observed.
\bc
\ec

\begin{figure}[htb]
\includegraphics[width=0.60\textwidth]{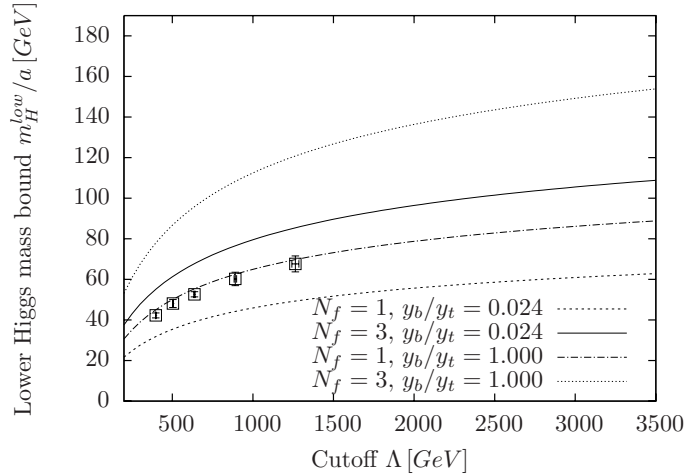} 
\caption{The various curves show the cutoff dependence of the lower Higgs boson mass bound in the infinite volume limit
derived from the effective potential for different physical setups ($N_f, y_b/y_t$). 
These analytical findings are compared with the infinite volume extrapolation of the Monte-Carlo data taken from
Tab.~\ref{tab:ResultOfExtrapolation} depicted by the square symbols. The numerical results were obtained
at ($N_f=1, y_b/y_t=1$) and are in very good agreement with the corresponding dash-dotted curve.
The solid curve represents the setup ($N_f=3, y_b/y_t=0.024$) coming closest to the actual physical situation in the
Standard Model. }
\label{fig:LowerBoundPrediction}
\vs{-2mm}
\end{figure}

Finally, we present the analytical infinite volume results, which are obtained from the effective potential in Eq.~(\ref{eq:ImprovedEffPot}) 
by replacing the sums with corresponding integrals. In Fig.~\ref{fig:LowerBoundPrediction} 
we compare these findings to the infinite volume extrapolation of the Monte-Carlo data taken from 
Tab.~\ref{tab:ResultOfExtrapolation} and observe very good agreement. We also show the analytical infinite volume
expectations for several other physical scenarios with varying values for $N_f$ and $y_b/y_t$ to study the influence of these parameters on the
lower mass bound. The number of fermion generations (or here equivalently the number of colours) as well as the bottom-top mass splitting
have a strong impact on the lower mass bound. The solid curve in Fig.~\ref{fig:LowerBoundPrediction} with $N_f=3$ and $y_b/y_t=0.024$ comes closest 
to the actual situation in the Standard Model. However, direct Monte-Carlo simulations in this scenario on sufficiently large lattices
would be too demanding with our available resources due to the huge mass splitting between the top and bottom quark.

\bc
\ec

\section{Outlook towards upper Higgs boson mass bounds}

\begin{figure}[htb]
\begin{tabular}{c}
\begin{tabular}{cc}
\includegraphics[width=0.48\textwidth]{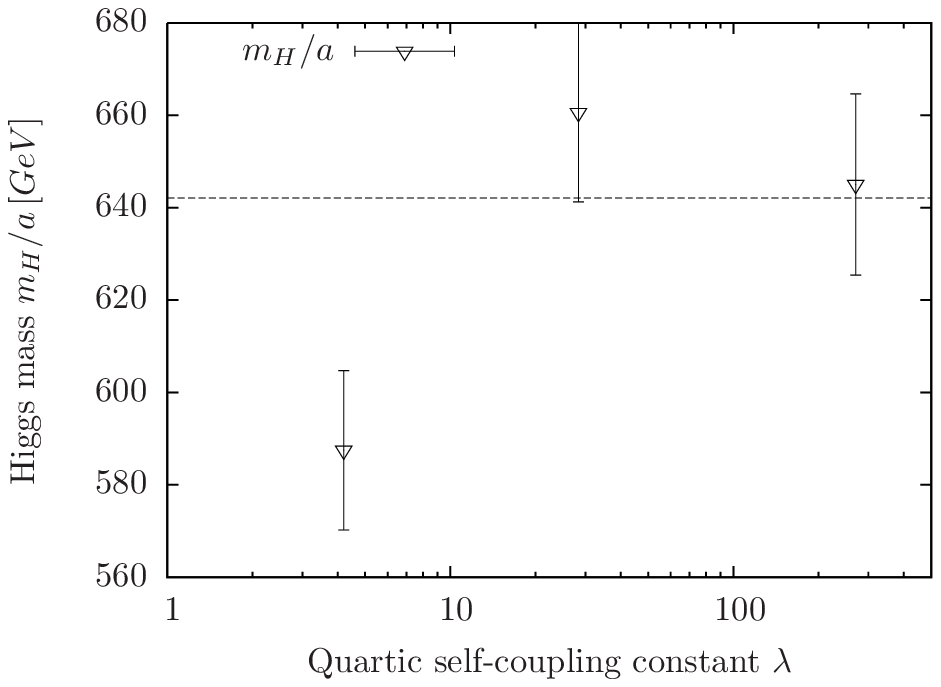}
&
\includegraphics[width=0.48\textwidth]{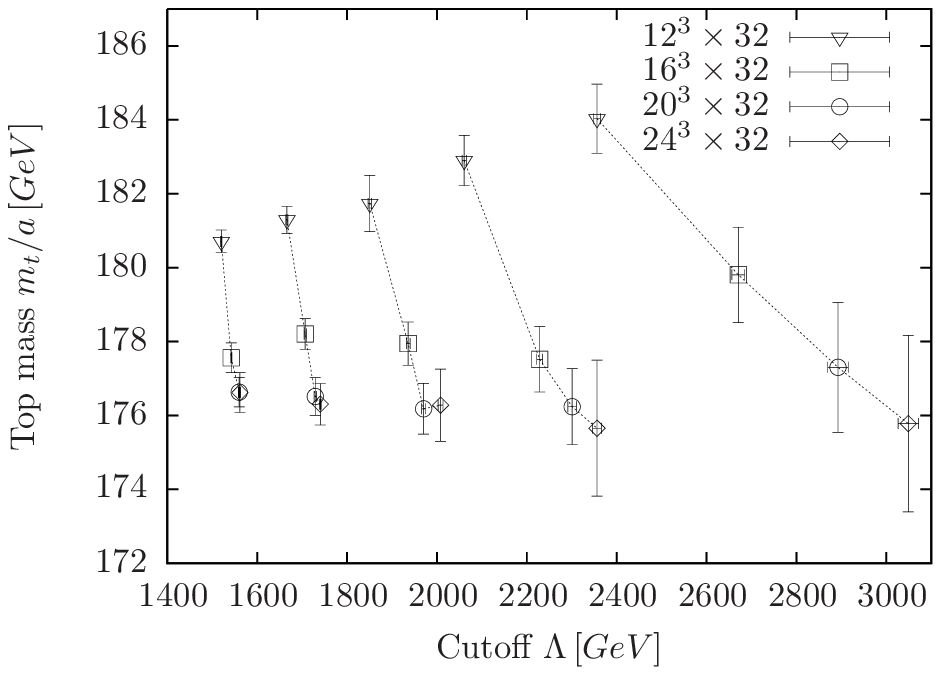}\\
\hs{4mm}(a) & \hs{8mm}(b)  \\
\end{tabular}\\
 \\
\includegraphics[width=0.60\textwidth]{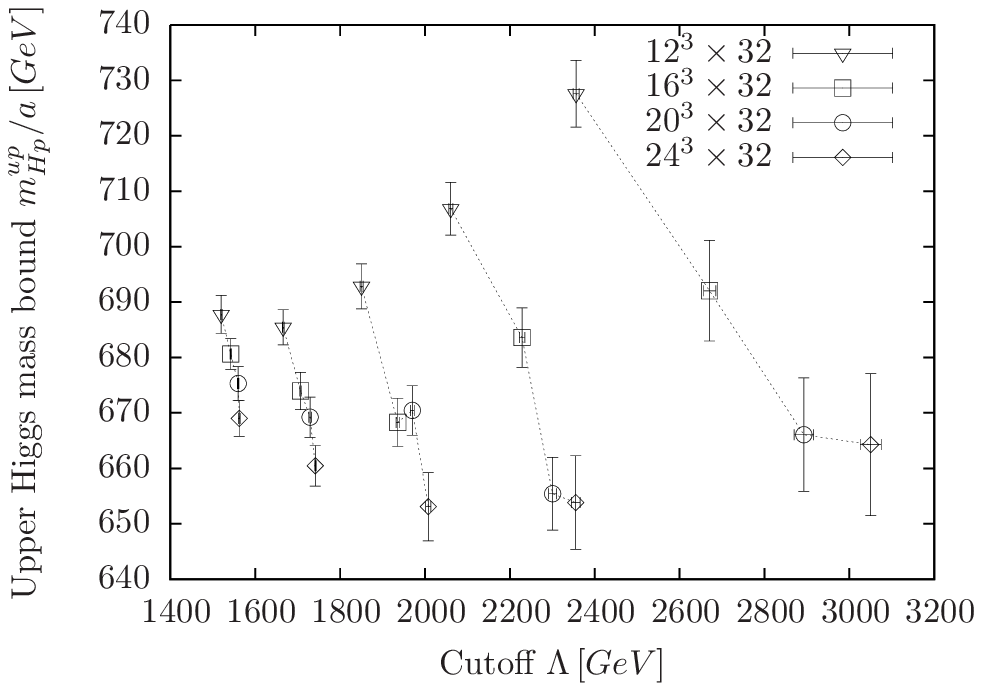}  \\
(c) \\
\end{tabular}
\caption{(a) The correlator Higgs boson mass versus the quartic coupling constant $\lambda$ in the
strong coupling region of the model at $\Lambda\approx 1500\,\mbox{GeV}$  on a $12^3\times 32$-lattice. 
The dashed line depicts the corresponding $\lambda=\infty$ result.
(b) The cutoff dependence of the top correlator mass at $\lambda=\infty$.
(c) The cutoff dependence of the Higgs propagator mass at $\lambda=\infty$.
Finite volume effects are illustrated in panels b, c by rerunning the simulations
on several lattice sizes. Runs with identical bare parameter settings are connected by lines to guide the eye.
All presented results have been obtained in Monte-Carlo simulations with identical, degenerate
bare Yukawa coupling constants fixed according to Eq.~(\ref{eq:treeLevelTopMass}) and $N_f=1$.
}
\label{fig:StrongLambdaDependence}
\vs{-2mm}
\end{figure}

We now turn towards the determination of the upper Higgs boson mass bound $m_H^{up}(\Lambda)$. 
First, we investigate whether the largest Higgs boson masses are obtained at $\lambda=\infty$ as expected from perturbation theory. 
This can indeed be observed in  Fig.~\ref{fig:StrongLambdaDependence}a where we plot the physical Higgs boson mass $m_H/a$ versus
the quartic self-coupling constant in the strong coupling region of the model for a fixed cutoff $\Lambda\approx 1500\,\mbox{GeV}$. 
We compare these numerical findings for large but finite $\lambda$ with the corresponding $\lambda=\infty$ result, which was 
obtained from our Monte-Carlo simulations by reparametrizing the Higgs field $\phi_x$ in terms
of $SU(2)$ elements. Since the integration of the molecular dynamics equations as needed in our Monte-Carlo 
algorithm becomes increasingly difficult
with growing $\lambda$, we present here only a few results on rather small $12^3\times 32$-lattices.
However, within the available accuracy one can conclude from Fig.~\ref{fig:StrongLambdaDependence}a that
the largest Higgs boson masses are obtained at $\lambda=\infty$, as expected. 
We therefore derive the upper Higgs boson mass bounds in the following from simulations with infinite self-coupling
constant. 

As discussed in the previous section the accessible energy scales $\Lambda$ are restricted
by Eq.~(\ref{eq:RequirementsForLatMass}). Employing a top mass of $175\,\mbox{GeV}$ and a Higgs boson 
mass of below $700\,\mbox{GeV}$, as suggested by Fig.~\ref{fig:StrongLambdaDependence}c, we should be 
able to reach cutoffs between $1400\,\mbox{GeV}$ and $2100\,\mbox{GeV}$ on a $24^3\times 32$-lattice. 

In Fig.~\ref{fig:StrongLambdaDependence}b we present the results for the top quark mass obtained in 
our Monte-Carlo calculations with degenerate bare Yukawa coupling constants fixed according to 
Eq.~(\ref{eq:treeLevelTopMass}) and $N_f=1$. Again we estimate the strength of the finite volume effects by 
rerunning the simulations on different lattice sizes. One sees that for the simulation runs with
$\Lambda < 2.1\,\mbox{TeV}$ the volume dependence of the top quark mass has already reached its saturation
plateau on the $24^3\times 32$-lattices within the given accuracy.
In comparison to the weak coupling region of the model the top quark mass depends significantly weaker on the cutoff. 
The discrepancy between the tree-level determination of the bare Yukawa coupling constants and the targeted top quark mass 
is around $1\%$ for the largest presented lattices and thus smaller than in the weak coupling region. However, in follow-up
calculations the available data can be used for a fine tuning of the bare Yukawa coupling constants to reproduce
the targeted fermion masses even more precisely.

Finally, we present our results on the cutoff dependence of the upper Higgs boson mass bound in Fig.~\ref{fig:StrongLambdaDependence}c.
Here, we decided to show only the propagator masses instead of the correlator masses.
The reason is that the latter results are very noisy, since they are
based on only those $L_t$ Higgs modes in momentum space which have spatial momentum zero,
and do - at the time being - not allow for a clear observation of the cutoff dependence 
of the Higgs boson mass. The propagator masses, on the other hand, are much more stable and do reveal the desired cutoff dependence
of the upper mass bound. On the small $12^3\times 32$-lattices the Higgs boson mass seems to increase with
the cutoff $\Lambda$. This, however, is only a finite volume effect. On the larger lattices the Higgs boson mass 
decreases with growing $\Lambda$ as expected. Only when the cutoff becomes too high for the given lattice volume does 
the Higgs boson mass finally start increasing again due to the finite volume. Larger lattice sizes and more statistics are needed
here to further follow the cutoff dependence of the upper mass bound.

\bc
\ec

\section{Summary and outlook}

The eventual aim of our work is to establish non-perturbative upper and lower Higgs boson mass
bounds based on first principle computations without relying on arguments such as 
vacuum instability or triviality. This can be achieved by investigating
a lattice model of the pure Higgs-Yukawa sector of the Standard Model.
The main idea of this approach is to apply direct Monte-Carlo simulations to determine the maximal 
interval of Higgs boson masses attainable within this model being consistent with phenomenology, 
\ie with the physical values of the top and bottom quark masses and the vacuum expectation value of the
Higgs mode. To maintain the chiral character of the Higgs-fermion coupling structure on the lattice 
we have considered here a chirally invariant lattice Higgs-Yukawa model based on the Neuberger Dirac operator.

In the present paper we confirmed that the lightest Higgs boson masses are generated at vanishing quartic self-coupling, as expected
from perturbation theory. For the mass degenerate case with equal top and bottom quark masses 
and $N_f=1$ we presented our numerical data for the Higgs boson mass obtained at $\lambda=0$ for various cutoffs 
and lattice sizes. These results were compared to the analytical calculations of the Higgs boson mass based on the effective potential
at one-loop order. We then extrapolated the numerical finite size results to infinite volume and compared them 
to the corresponding infinite volume effective potential predictions. In both cases very good agreement was 
observed. 

The main result of the presented findings is that a lower Higgs boson mass bound is a manifest property of the pure
Higgs-Yukawa sector that evolves directly from the Higgs-fermion interaction for a given Yukawa coupling parameter.
For growing cutoff this lower mass constraint rises monotonically with flattening slope as expected
from perturbation theory. Moreover, the quantitative size of the lower bound is comparable to the magnitude of the
perturbative results based on vacuum stability considerations~\cite{Hagiwara:2002fs}. A direct quantitative comparison
is, however, non-trivial due to the different regularization schemes accompanied with the strong cutoff dependence
of the considered constraints.

We also studied the Higgs boson mass dependence on the top-bottom mass splitting on some smaller lattices,
where the calculations were feasible. Again the numerical findings matched the corresponding results 
based on the effective potential. The very good agreement between the numerical and analytical approaches justifies
the presented effective potential calculations performed in the actual physical 
situation with $N_f=3$ and $m_t/m_b\approx 40$. A comparison with direct Monte-Carlo simulations 
in that setup on sufficiently large lattice sizes, however, seems to be too demanding for our available resources at the moment.
We therefore have to rely on the analytical results in this case for the time being.

Additionally, we turned towards the upper Higgs boson mass bound, which was obtained at infinite bare self-coupling
after having checked that the heaviest Higgs boson masses are indeed generated at $\lambda=\infty$ as expected.
Instead of the correlator masses we decided to present the propagator masses at this point, since the latter
results were stable enough to allow for the observation of the desired cutoff dependence of the upper Higgs boson 
mass bound in contrast to the correlator masses. Apart from finite volume effects the expected 
decrease of the upper bound with growing cutoff could clearly be seen.

As a next step we will improve the statistics for the upper Higgs boson mass bound, extend the lattice volumes, 
and compare the numerical findings to corresponding calculations in renormalized perturbation theory,
once sufficient precision is achieved. For the lower bound we plan to include higher order terms into the 
bare Lagrangian in order to investigate the universality of the obtained results. 
Taking these future improvements aside, our present results suggest a maximally allowed
Higgs boson mass interval of $80\mathrm{Gev} \lesssim m_H \lesssim 700\mathrm{Gev}$ at a cutoff around $\Lambda\approx 1\,\mbox{TeV}$.
Eventually, we are also interested in studying the decay properties of the Higgs boson with this lattice model.

\section*{Acknowledgments}
We thank J. Kallarackal for discussions and M. M\"uller-Preussker for his continuous support.
We are grateful to the "Deutsche Telekom Stiftung" for supporting this study by providing a Ph.D. scholarship for
P.G. We further acknowledge the support of the DFG through the DFG-project {\it Mu932/4-1}.
The numerical computations have been performed on the {\it HP XC4000 System}
at the {\it Scientific Supercomputing Center Karlsruhe} and on the
{\it SGI system HLRN-II} at the {\it HLRN Supercomputing Service Berlin-Hannover}.

\bibliographystyle{unsrtOWN}
\bibliography{HiggsYukawaMassBoundsFirstResults}

\end{document}